\documentclass[prl,twocolumn,superscriptaddress,floatfix,preprintnumbers,amssymb,amsmath]{revtex4-2}

\usepackage{hyperref}
\usepackage{graphicx}
\usepackage{xcolor}
\usepackage{ulem}

\begin{document}
\title{On chip synchronization of Bloch oscillations in a strongly coupled pair of small Josephson junctions}

\author{Fabian~Kaap}
\affiliation{Physikalisch-Technische Bundesanstalt, Bundesallee 100, 38116 Braunschweig,
	Germany}
\author{David~Scheer}
\affiliation{JARA Institute for Quantum Information, RWTH Aachen University, 52056 Aachen, Germany}
\author{Fabian~Hassler}
\affiliation{JARA Institute for Quantum Information, RWTH Aachen University, 52056 Aachen, Germany}
\author{Sergey~Lotkhov}
\email{Sergey.Lotkhov@ptb.de}
\affiliation{Physikalisch-Technische Bundesanstalt, Bundesallee 100, 38116 Braunschweig,
Germany}

\begin{abstract}
Bloch oscillations are a fundamental phenomenon linking the adiabatic transport of Cooper pairs to time. Here, we investigate synchronization of the Bloch oscillations in a strongly coupled system of sub-100 nm Al/AlO$_{\rm x}$/Al Josephson junctions in high-ohmic environment composed of highly inductive meanders of granulated aluminum and high-ohmic titanium microstrips. We observe a pronounced current mirror effect in the coupled junctions and demonstrate current plateaus, akin to the first dual Shapiro step in microwave experiments. These findings suggest that our circuit design holds promise for realizing protected Bloch oscillations and precise Shapiro steps of interest for current metrology.
\end{abstract}
\maketitle

Well-controllable, adiabatic charge transport in small Josephson junctions is a promising avenue for metrological applications \cite{LikharevZorin1985}. In particular, in the regime of Bloch oscillations, the transported current $I_\text{B}$ is fundamentally linked  to the oscillation frequency $f_\text{B}$ via $I_{\rm B} = 2e \times f_{\rm B}$, in the nA respectively GHz ranges of practical interest. However, the small width of the lowest Bloch band $\delta E_{\rm 0}$, k$_{\rm B}T_{\rm eff} \alt \delta E_{\rm 0} < E_{\rm C} \sim E_{\rm J}$, makes these oscillations  challenging to directly measure.  Here, $E_{\rm C} = e^2/2C \sim E_{\rm J} = \hbar I_{\rm c}/2e \sim100\,\mu$eV are the charging and the Josephson energies, respectively, for a typical Al junction of capacitance $C \sim 1\,$fF and critical current $I_{\rm c} \sim50\,$nA, thermalized at an effective electron temperature $T_{\rm eff} \sim 100\,$mK. 

In a number of experiments (see, $e.g.$, Refs.~\cite{KuzminPhysicaB,Shaikhaidarov2022, Crescini2022}), Bloch oscillations were synchronized to an external  drive at fixed frequency. This gives rise to so-called dual Shapiro steps of constant current predicted by theory \cite{LikharevZorin1985,MooijNazarov2006}. The Bloch oscillations demand for high-ohmic biasing resistors, implemented in order to reduce the quantum fluctuations of charge and to suppress inelastic tunnel processes. In early experiments, this resulted in strong electron overheating, the dramatic increase of thermal fluctuations, and, finally, in washing out the coherent oscillation effects \cite{KuzminPhysicaB,CorleviPRL2006,MaibaumPRB2011}. 

As a remedy, a combined inductively-resistive environment was  theoretically proposed as a viable alternative to the purely resistive damping \cite{Arndt2018}. This alternative approach has recently been implemented using either high-kinetic inductance elements made of thin NbN films in Ref.~\cite{Shaikhaidarov2022} or Josephson inductances in a long array of classical (large area) Josephson junctions \cite{Crescini2022}. However, as shown in Ref.~\cite{Shaikhaidarov2022}, inelastic tunneling processes, that come into play due to finite effective temperature and high, but limited bias impedance, still impose severe restrictions on the flatness of the resulting current steps.  

\begin{figure}[tb]
	\centering\includegraphics[width=\columnwidth]{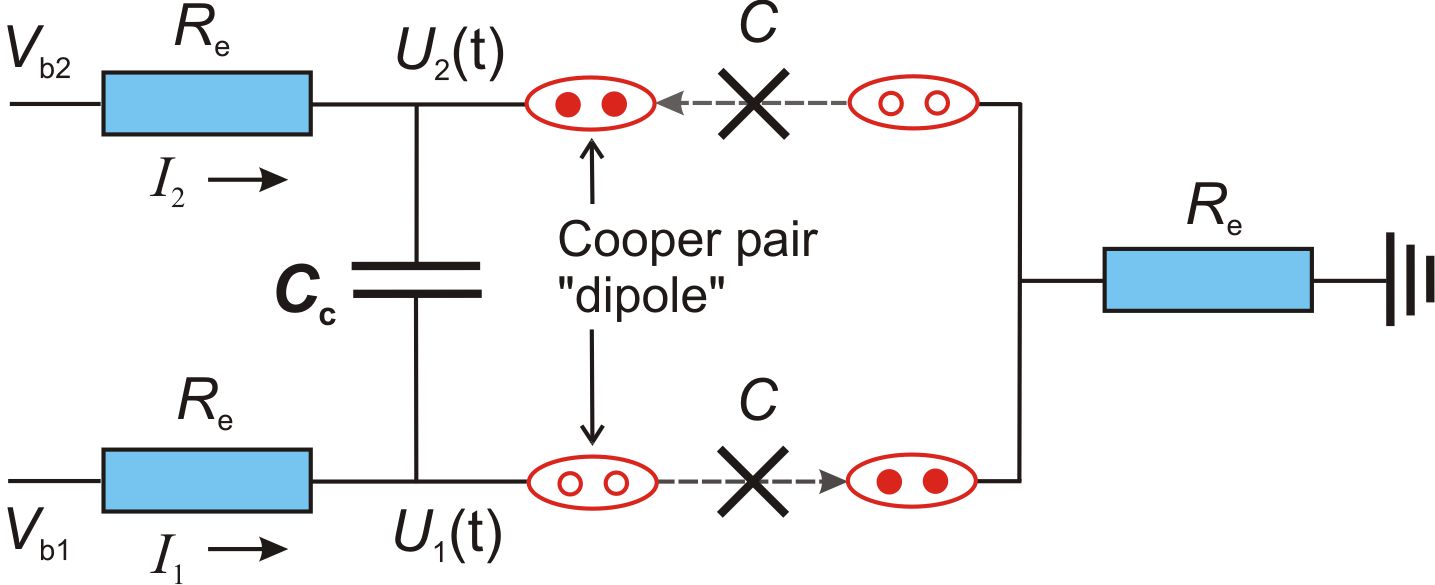}
	\caption{(Color online) A symmetric circuit with two coupled Josephson junctions allowing for the synchronization of Bloch oscillations. The signature of this is the current mirror effect with $I_1 = - I_2$, appearing due to transport via Cooper pair `dipoles' indicated in red, see main text for details. Note that in order to achieve coherent Bloch oscillations, the bias resistances must exceed the resistance quantum with   $R_{\rm e} \gg R_{\rm Q} = h/4e^2 \approx6.45\,$k$\Omega$.}
	\label{Fig1}
\end{figure}

Considerable improvement could be achieved using the dipole charge transfer in a symmetric, strongly-coupled pair of Bloch junctions, see Fig.~\ref{Fig1}, with different energy scales for coherent, $\sim e^2/2C_{\rm c}$, and inelastic, $\sim E_{\rm C}$, Copper pair processes, where $C_{\rm c} \gg C$ is the coupling capacitance. In such a circuit, the competing single Cooper pair processes are expected to be subdominant compared to the coherent dipole transfers keeping the voltage across the junctions small, $U_{\rm 1,2} \sim (e/C_{\rm c})/2 \ll e/C$. A related setup has been proposed in Ref.~\cite{Kitaev2006} for constructing a protected qubit based on dipole transport (dubbed `current mirror') in a Josephson ladder. Experimentally, the current mirror effect with the signal correlation of up to 98.8$\%$ was demonstrated by studying the switching voltage statistics in two superconducting 1D-arrays \cite{Shimada2000}. 

\begin{figure}[tb]
	\centering\includegraphics[width=\columnwidth]{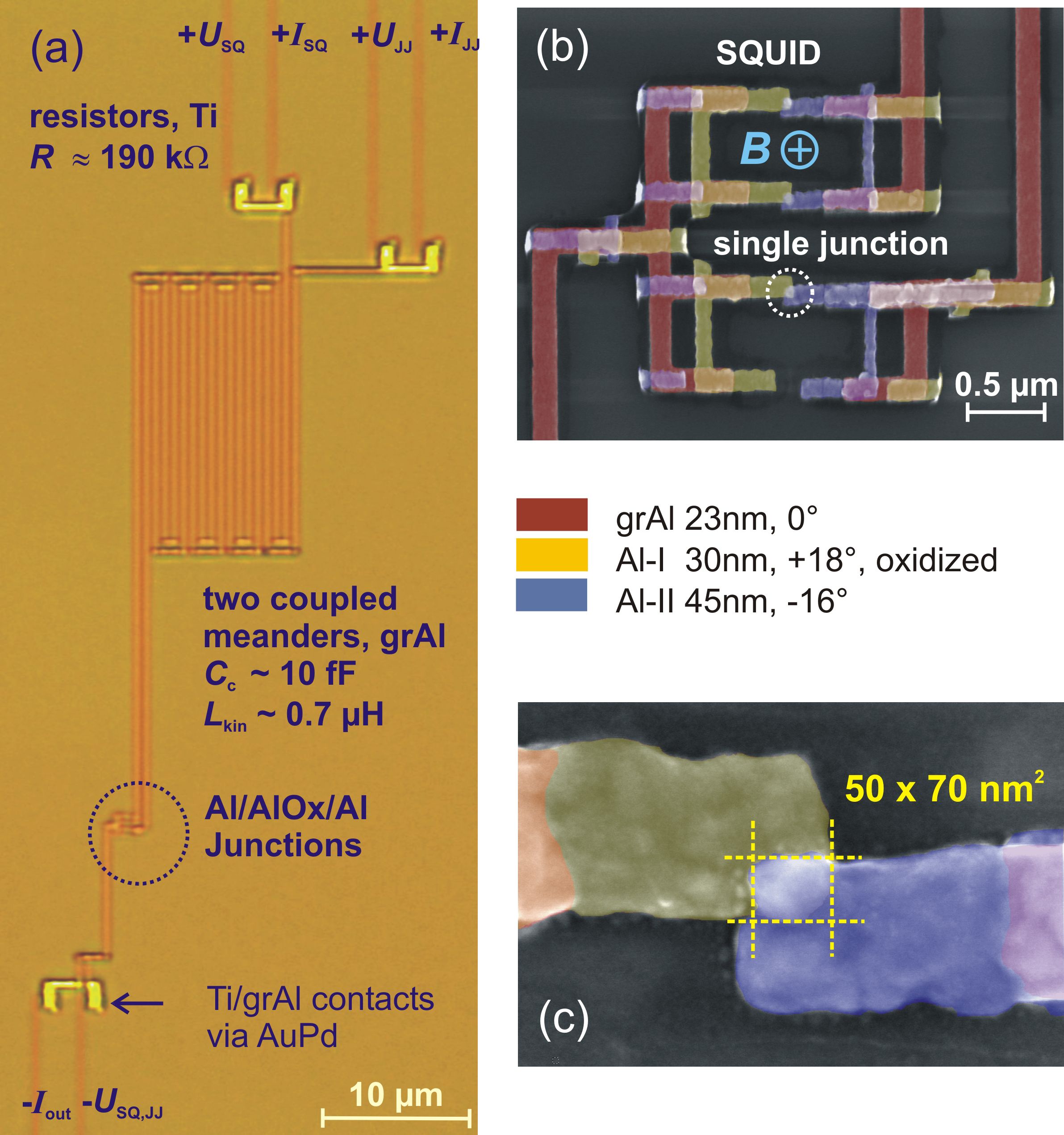}
	\caption{(Color online) (a) Optical picture of high-impedance circuitry. Both current branches include titanium microstrips, 20$\times$0.16$\times$0.15 $\mu$m$^3$ long/wide/thick each, as well as 0.14$\,\mu$m wide meandered  lines of  grAl. (b) False-colored SEM image of the SQUID and single junction structure. Comparing to Fig.~\ref{Fig1}, the junction (SQUID) plays the role of the first (second) branch.  (c) Close-up view of a single junction marked by the dashed circle in (b).}
	\label{Fig2}
\end{figure}

In this Letter, we present the results of our DC characterization for a strongly-coupled pair of Bloch oscillators enabling the dipole transport discussed above. We study an asymmetric tunable structure including a small SQUID coupled to a single Bloch junction, shown in Figs.~\ref{Fig2}(a,b). The junctions were biased via two capacitively coupled meanders in series to high-ohmic resistive leads. We show the tunable-capacitor behavior, observed at zero bias current in SQUID and also, more importantly, the current mirror effect in the two-currents plane resulting from the synchronization of the two Bloch oscillations in the sub-GHz range. Due to the synchronization, the data  exhibit a pronounced current step with a step value that is akin to the first dual Shapiro step induced by an external GHz signal \cite{KuzminPhysicaB,Shaikhaidarov2022, Crescini2022}. The data are consistent with a theory providing an interpretation in terms of correlated adiabatic ground state transport.

The experimental circuit was fabricated using e-beam-exposed PMMA/copolymer liftoff masks with a controllable undercut \cite{LecocqNanotechn2018} and the shadow evaporation technique \cite{DolanAPL1977}. The Josephson junctions of type Al/AlO$_{\rm x}$/Al were deposited in the common three-angle evaporation process \cite{KaapCPEM2022} with the coupling meanders made of granulated aluminum (grAl), see, $ e.g. $, Ref.~\cite{GruenhauptThesis2019}. The first Al layer was oxidized at $P_{\rm O2}=0.3\,$Pa for 5\,min to enable small junctions, see Fig.~\ref{Fig2}(c), of low normal resistance, $R_{\rm N} \approx 6\,$~k$\Omega$, and high values of $E_{\rm J} \approx 110\,\mu$eV and  $E_{\rm C} \approx 150\,\mu$eV, estimated roughly for a single junction using co-fabricated test devices \cite{KaapCPEM2022}. We note that a similar device with slightly different parameters exhibited widely identical behavior as reported below.

For high-impedance biasing, the meander geometry was preferred, in order to minimize the stray lead capacitance \cite{KamenovPRA2020}. The kinetic inductance of grAl was estimated within the BCS approximation \cite{RotzingerSupercScieTechn2017}, to achieve $L_{\rm kin,\square} = 0.18\hbar R_{\rm n}/k_{\rm B} T_{\rm c} \approx 0.5\,$nH, for the measured values of the grAl normal state resistance, $R_{\rm n,\square} \approx700\,\Omega$, and $T_{\rm c} \approx1.9\,$K. As a result, the resonance frequency of the TEM wave along the meanders,  $\omega_{\rm p} = (2L_{\rm kin}C_{\rm c})^{-1/2}\sim 2\pi\times1.3\,$GHz, ranged over the frequencies in our experiment. At lower frequencies, the condition $R_{\rm e} \gg R_{\rm Q}$ was ensured by high-ohmic titanium microstrips co-evaporated with oxygen \cite{LotkhovNanotechn2013} at $P_{\rm O2}=3\times 10^{\rm -4}\,$Pa and featured by high film resistivity, $r \approx1.5\,$k$\Omega$ per square.

The sample was measured in a filtered environment at  $T \approx 15\,$mK. The DC currents were supplied symmetrically through pairs of room temperature resistors, $R_{\rm b}= 2\times 100\,$M$\Omega$, reaching the sample via $\ell \approx 1\,$m long Thermocoax$\textsuperscript{TM}$ filters of capacitance $c \ell\sim0.5\,$nF each \cite{ZorinRSI1999}, mounted in direct vicinity  of the chip. At microwave frequencies, such an environment realizes the condition of voltage biasing adopted in Fig.~\ref{Fig1}.

\begin{figure}[tb]
	\centering\includegraphics[width=\columnwidth]{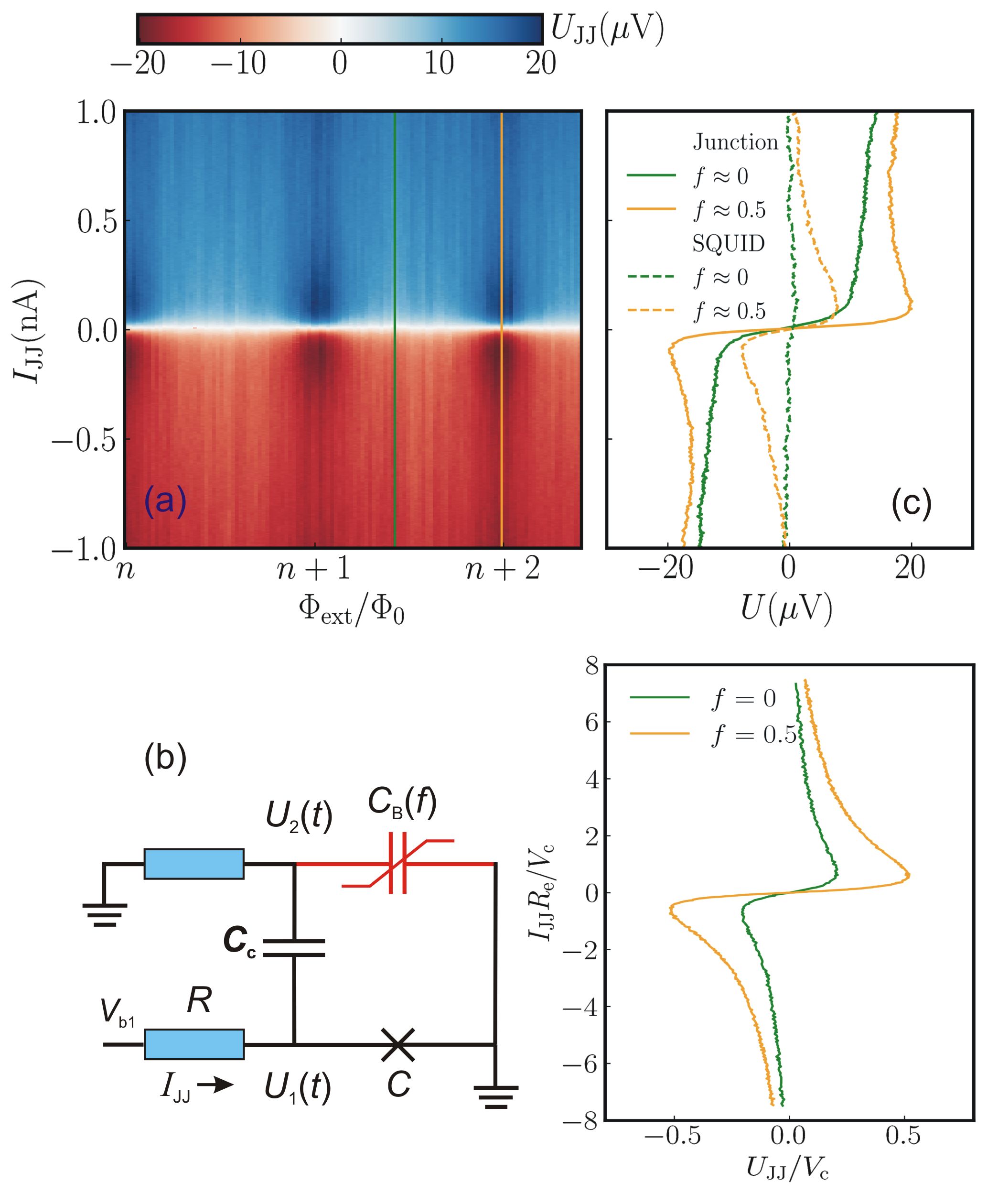}
	\caption{(Color online) (a) DC junction voltage, $U_{\rm JJ} = \overline{ U_1(t)}$, measured as a function of current $I_{\rm JJ} $ and magnetic flux $\Phi_{\rm ext}$ at $I_{\rm SQ} =0$. The green (orange) vertical line marks weak (strong) magnetic frustration of the SQUID, for which the cross-sectional $IV$-curves are plotted in (c). The dashed lines in (c) depict the DC voltage $U_{\rm SQ} = \overline{U_2(t)}$ induced across the SQUID. (b) Equivalent circuit as a single junction shunted via the Bloch capacitance realized by the SQUID. (d) Simulation of the $IV$-curves. The values of the parameters are those of Figs.~\ref{Fig4}(c) and (d)}
	\label{Fig3}
\end{figure}

The diagram in Fig.~\ref{Fig3}(a) maps the voltage $U_{\rm JJ}$ across the junction vs.\ bias current $I_{\rm JJ}$ at zero current biasing of the SQUID. The magnetic field $B$ and therefore the loop frustration, $ f \equiv \Phi_{\rm ext} /\Phi_0$, are varied over a few modulation cycles. Here $\Phi_{\rm ext} = B A$ is the flux threading the loop area, $A \approx 0.4\,\mu$m$^2$ [see SEM image in Fig.~\ref{Fig2}(b)], and $\Phi_0 = h/2e$ the magnetic flux quantum. In this small signal regime, illustrated in Figs.~\ref{Fig3}(b) and(c), the SQUID plays the role of a tunable Bloch capacitance $C_{\rm B}(U_{\rm 2},f) \approx C_{\rm B}(0,f)$ (see, $e.g.$, Ref.~\cite{Buettiker1987}), in series to the inter-meander capacitance $C_{\rm c}$ and both in parallel to the single junction \cite{kinetic}. At weak frustration, $f \approx0 \mod1$ ($B=23.1\,$mT), with a strong Josephson coupling in the SQUID, the junction turns out to be effectively shunted by a larger capacitance $C_{\rm c}$: $C \ll C_{\rm c} \ll C_{\rm B}(0,0)$. This leads to a small Coulomb blockade threshold, see Figs.~\ref{Fig3}(c) and (d). On the contrary, at strong frustration, $f \approx 0.5\mod1$ ($B=25.9\,$mT), the Bloch and therefore the total shunting capacitance are reduced, resulting in a stronger Coulomb blockade effect. The $IV$-curve shows a clear back-bending, a `fingerprint' of Bloch oscillations.

\begin{figure}[tb]
	\centering\includegraphics[width=\columnwidth]{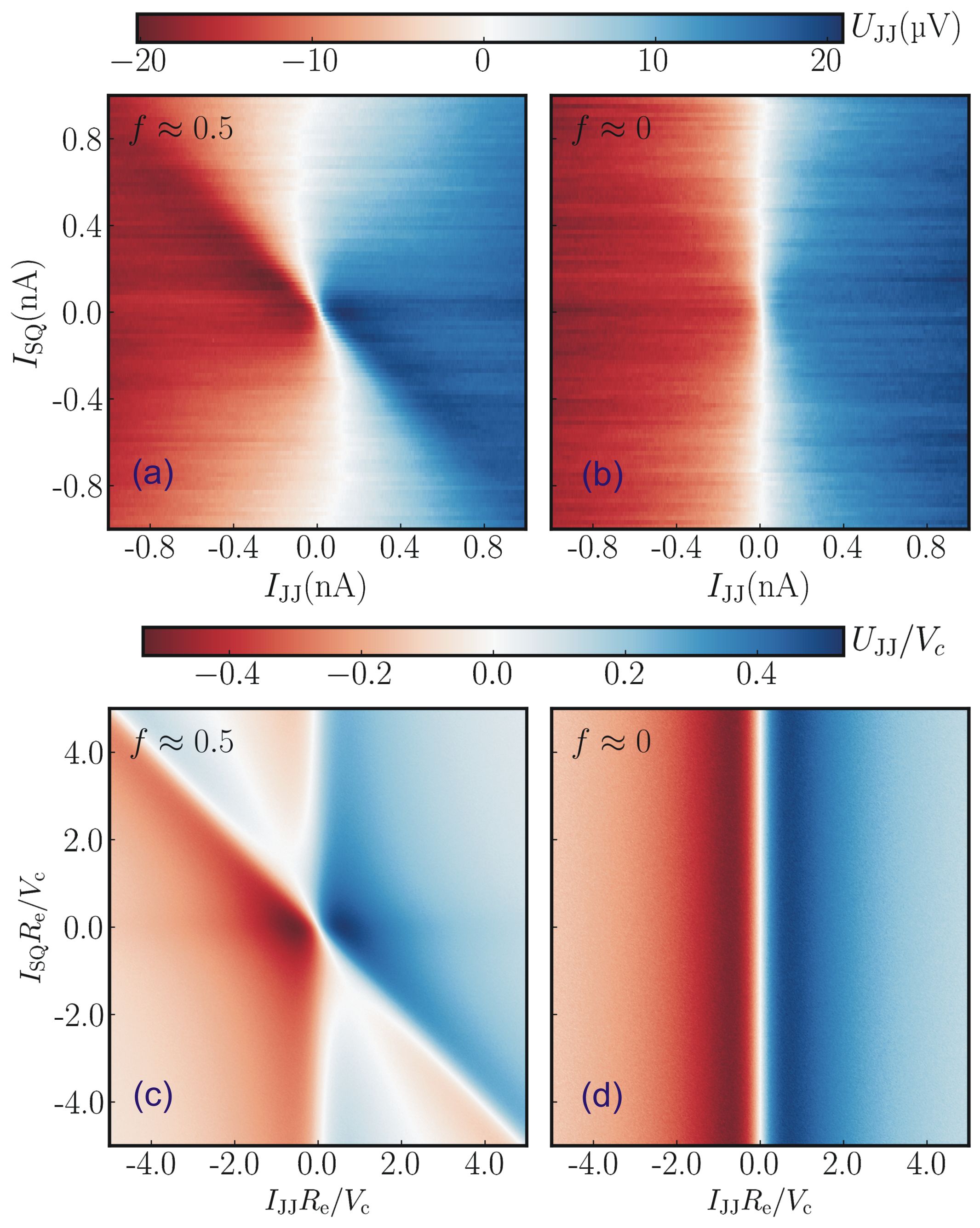}
	\caption{(Color online) (a,b) Voltage $U_{\rm JJ}$ as a function of the two bias currents for the two frustration cases.  (c) and (d) show the simulated data with Johnson-Nyquist noise at a constant temperature of $ k_{\rm B}T=0.25 eV_{{\rm c}1}$ in the regime of (c) symmetric Josephson energies and Bloch critical voltages, $V_{\rm c} \equiv V_{{\rm c}1}=V_{{\rm c}2}=V_+$, and (d) the single junction regime $V_{{\rm c}2}=V_+=0$. }	
	\label{Fig4}
\end{figure}

\begin{figure}[tb]
	\centering\includegraphics[width=\columnwidth]{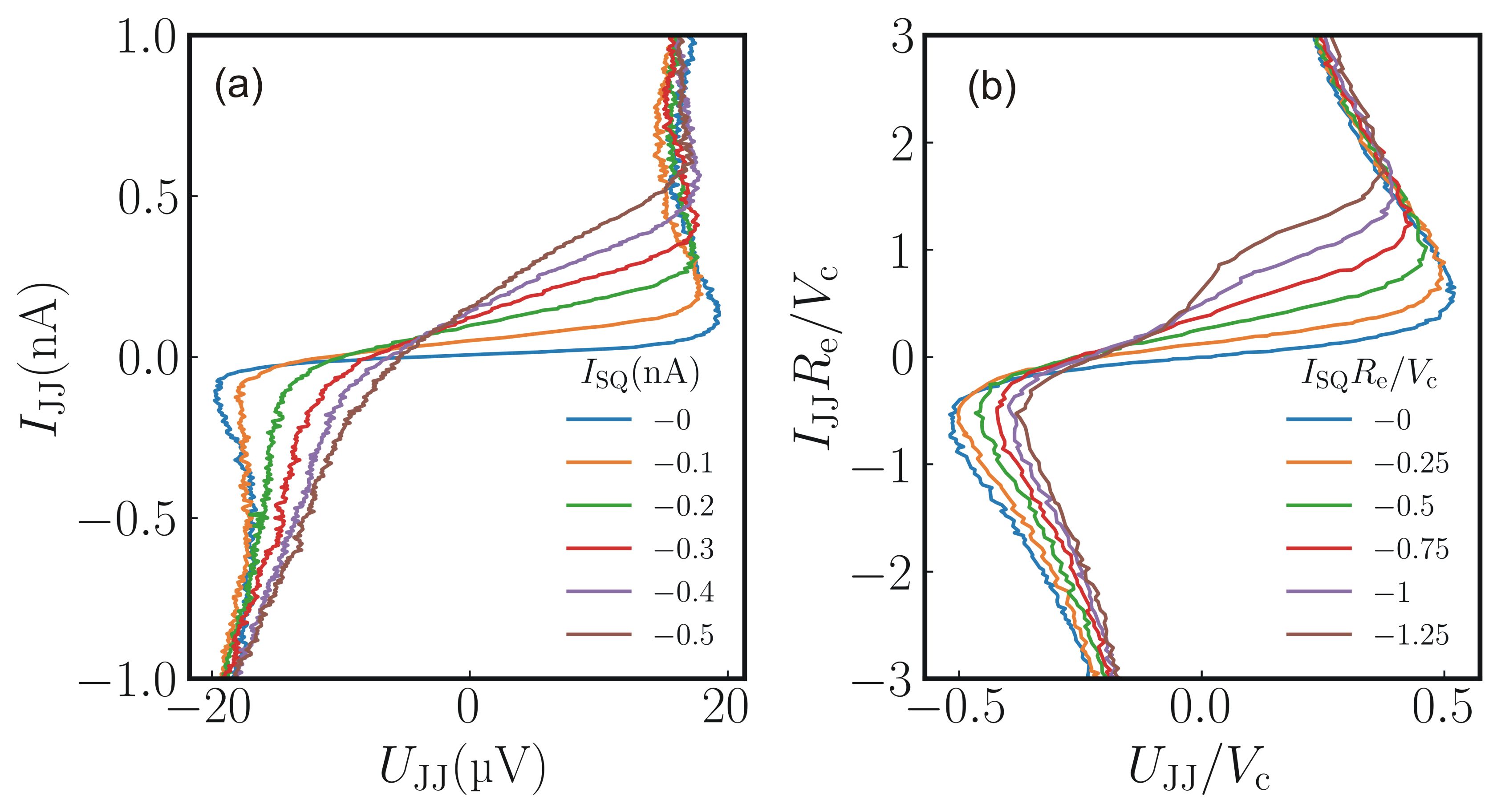}
	\caption{(Color online) (a) Current in the junction plotted  vs.\ junction voltage at $f \approx 0.5$ and $I_{\rm SQ}$ varying from 0 to 0.5\,nA (bottom to top). (b) Simulated data for the synchronized Bloch oscillations in the regime of symmetric Josephson energies for an equidistant series of SQUID current values $I_{\rm SQ}$.}
	\label{Fig5}
\end{figure}

In the two-current plane $(I_{\rm JJ},I_{\rm SQ})$, shown in Fig.~\ref{Fig4}, the voltage landscape also depends on the magnetic frustration in the SQUID. The diagram in Fig.~\ref{Fig4}(a), measured close to a half-integer value of $f$, is asymmetric and shows a singularity of the profile along the line $I_{\rm JJ} = - I_{\rm SQ}$, demonstrating the current mirror effect in the circuit. Qualitatively, the same voltage profile has been observed for the SQUID voltage $U_{\rm SQ}$ as well (not shown). In this regime, the Bloch oscillations in both arms are of appreciable amplitude. They are coupled to each other via the capacitance $C_{\rm c}$, giving rise to the plateaus with slow current ramping, $I_{\rm JJ}(U_{\rm JJ})$, shown in Fig.~\ref{Fig5}(a). The ramping of the plateaus, weak at small currents,  $I_{\rm SQ} \sim0.1\,$nA, is increased towards higher drive currents, in a good agreement with the simulated dynamics, the results shown in Fig.~\ref{Fig5}(b). Our numerical analysis described in detail below shows that  the plateaus are due to synchronization, akin to the first dual Shapiro step reported for the microwave drive signals in Refs.~\cite{KuzminPhysicaB,Shaikhaidarov2022, Crescini2022}. 
 
On the contrary, at weak frustration and strong SQUID coupling, we measure a smooth and symmetric disappearance of the blockade feature, as the absolute value of current $I_{\rm SQ}$ is increasing, see Fig.~\ref{Fig4}(b). This behavior, typical for growing thermal fluctuation effects, is caused by electron overheating in the titanium resistors \cite{Tat100pA}. The current mirror effect due to the synchronization is missing [cf. Fig.~\ref{Fig4}(d)] as the amplitude of the Bloch oscillations in SQUID is small [cf., $e.g.$, Fig.~\ref{Fig3}(c)].

We compare the experimental data with the results of simulations based on Kirchhoff's equations  
\begin{equation*}
	V_{{\rm b}i}=R_\text{e} (\dot Q_i+  \dot Q_+) +V_{{\rm c}i}\sin\Bigl(\frac{\pi
		Q_i}{e}\Bigr)+V_+\sin\Bigl(\frac{\pi Q_+}{e}\Bigr),
\end{equation*} 
for the charges $Q_i$ supplied by the individual leads with a combined charge $Q_+=Q_1+Q_2$. In addition to the bias voltages $V_{{\rm b} i}$ and the voltage drops over the resistors $R_\text{e}$ (first term), the equations contain the voltage drops $U_i$ across the junctions (remaining terms), see Ref.~\cite{Supplement} for the derivation. It consists of the voltage due to individual phase slips with a critical voltage $V_{{\rm c}i}$ and the term with critical voltage $V_+$ describing coupled phase slips due to $C_c$. It is this last term that synchronizes the Bloch oscillations and leads to the current plateaus.

We integrate the equations of motion using the forward Euler method to obtain the voltage drop $U_1(t)$ across the junction for different junction currents. In order to resolve the dynamics at the characteristic frequency of the junction's Bloch oscillations, $\omega_{R}=\pi V_{{\rm c}1}/eR_\text{e}$, we choose discrete time steps of the duration $\delta= 0.25/\omega_{R}$. We include the effects of Johnson-Nyquist noise from the biasing resistors as Gaussian white noise increments with standard deviation $2R_\text{e} k_{\rm B}T/\delta^{1/2}$. For simplicity and deviating from the experimental conditions with growing thermal smearing effects at increasing currents $I_{\mathrm{JJ}}$ and $I_{\mathrm{SQ}}$, the simulations were performed at a constant temperature $k_{\rm B} T= 0.25 eV_{{\rm c}1}$, which corresponds to $T \approx 150\,$mK at $V_{{\rm c}1}\approx50\,\mu$V.

We obtained simulation results for both strong and weak frustration of the SQUID. In the regime of strong frustration, $f \approx 0.5\mod 1$, the Josephson energies of the SQUID and the junction are almost equal. Under assumption of equal effective junction capacitances, this leads to comparable critical voltages,  $V_{{\rm c}1} \approx V_{{\rm c}2}$, with Bloch oscillations of similar strength. Due to the large coupling capacitance $C_{\rm c}$, the synchronization of the Bloch oscillations has a substantial contribution to the overall voltage drop with $V_+\approx V_{{\rm c}i}$, see Ref.~\cite{Supplement}. At weak frustration,  $f \approx 0\mod 1$, the SQUID branch becomes almost superconducting, see Fig.~\ref{Fig3} (c), with a large Bloch capacitance, $C_B \gg C_c$, shunting the SQUID. As a result, it exhibits only weak Bloch oscillations effects with  $V_{{\rm c}2}\approx V_+\approx 0$. 

Figure~\ref{Fig4}(c) shows the regime of strong magnetic frustration corresponding to symmetric Josephson energies. The simulations show a synchronization of Bloch oscillations leading to a current mirror effect, $I_{\mathrm{JJ}}=-I_{\mathrm{SQ}}$, whereas the synchronization leading to aligned currents is suppressed by thermal effects~\cite{Supplement}. Figure~\ref{Fig4}(d) shows the regime of weak magnetic frustration corresponding to Bloch oscillations only in the single junction. The simulation points to the absence of synchronization, in this case, as well as a diminished blockade threshold also shown in Fig.~\ref{Fig3}(d). 

\begin{figure}[tb]
	\centering\includegraphics[]{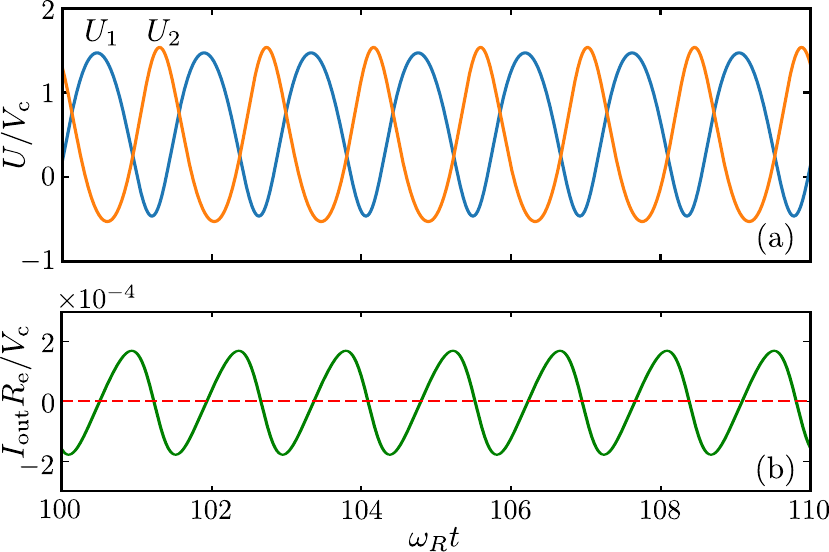}
	\caption{(Color online) Simulation of the time-resolved synchronization of Cooper pair transfers in the regime of symmetric Josephson energies with asymmetric bias voltages $V_{{\rm b}1}=5V_{{\rm c}1}$, $V_{{\rm b}2}=-4V_{{\rm c}1}$. (a) Voltage drop $U_1$ ($U_2$) across the junction (SQUID). (b) A small common node current $I_{\rm out}$ through the lower biasing resistor indicates a current mirror effect.}
	\label{Fig6}
\end{figure}

Due to the strong junction-to-SQUID coupling, $C_{\rm c} > C$, the current $I_{\mathrm{JJ}}$ exhibits plateaus at the level scaling with the current $-I_{\rm SQ}$ in the SQUID, as shown in Fig.~\ref{Fig5}(b). These plateaus emphasize the current mirror effect based on synchronization of Bloch oscillations. The synchronization can also be explicitly tracked in the time domain for the voltage drops $U_1(t)$ and $U_2(t)$ across the junction and SQUID, respectively. Figure~\ref{Fig6}(a) shows simulation results without noise, resolving oscillations on the scale of the inverse characteristic frequency. They indicate anti-correlated (or dipole-type) Cooper pair transfer in the two junctions, thus manifesting a current mirror effect on the microscopic scale. We note that the oscillation frequencies are matched even for asymmetric bias voltages, resulting in a small oscillating output current, $I_{\rm out} = I_{\mathrm{JJ}}+I_{\mathrm{SQ}}$, with an average close to zero, as shown in Fig.~\ref{Fig6}(b). 

The simulations exhibit good agreement with the experimental data for small bias currents, $I_{\rm JJ,SQ} \alt 0.5\,$nA $\ll \pi e E_\text{J}^2/4\hbar E_\text{C}\approx 16\,$nA, with negligible rate of Landau-Zener tunneling  (see, $e.g.$, Ref.~\cite{Geigenmueller88}). At higher currents, the Bloch oscillations effects vanish behind the inelastic Cooper pair transport, which shows up as softening of the `back-bending' in the measured $IV$-curves in Figs.~\ref{Fig3}(c) and \ref{Fig5}(a) as compared to Figs.~\ref{Fig3}(d)] and \ref{Fig5}(b), respectively.

To conclude, we have shown synchronization of coherent Bloch oscillations in a coupled circuit with ultrasmall Josephson junctions. In particular, we observed a current mirror effect and related plateaus of constant current in the DC $IV$ curves of the junctions. We have supported our interpretation of the data by theoretical analysis showing good qualitative agreement in a sub-GHz range, where the synchronization effect is pronounced. The demonstrated current plateaus are akin to the first dual Shapiro step, which paves the way towards the development of a quantum current standard. In the future, it would be interesting to optimize the current plateaus. In particular, the numerical simulations indicate that increasing $C_\text{c}$ will further stabilize the synchronization and thus the current mirror effect.

Fruitful discussions with A.~Zorin and L.~Grünhaupt are appreciated. The authors acknowledge technical support from P.~Hinze and T.~Weimann.
This work was supported by the Deutsche Forschungsgemeinschaft (DFG) under
Grant No.~HA 7084/7--1 and LO 870/2--1.


\begin{thebibliography}{23}
\bibitem{kinetic} Our simulations show that for Bloch currents below  $e \omega_\text{p} / \pi \sim 0.4\,$nA, a lumped-element, purely capacitive approximation neglecting kinetic inductance term appears to be qualitatively adequate.
\end{thebibliography}
\end{document}


\title{On chip synchronization of Bloch oscillations in a strongly coupled pair of small Josephson junctions}

\author{Fabian~Kaap}
\affiliation{Physikalisch-Technische Bundesanstalt, Bundesallee 100, 38116 Braunschweig,
	Germany}
\author{David~Scheer}
\affiliation{JARA Institute for Quantum Information, RWTH Aachen University, 52056 Aachen, Germany}
\author{Fabian~Hassler}
\affiliation{JARA Institute for Quantum Information, RWTH Aachen University, 52056 Aachen, Germany}
\author{Sergey~Lotkhov}
\affiliation{Physikalisch-Technische Bundesanstalt, Bundesallee 100, 38116 Braunschweig,
Germany}

\maketitle


\section*{Supplemental Material}

\begin{figure}[b]
	\centering\includegraphics{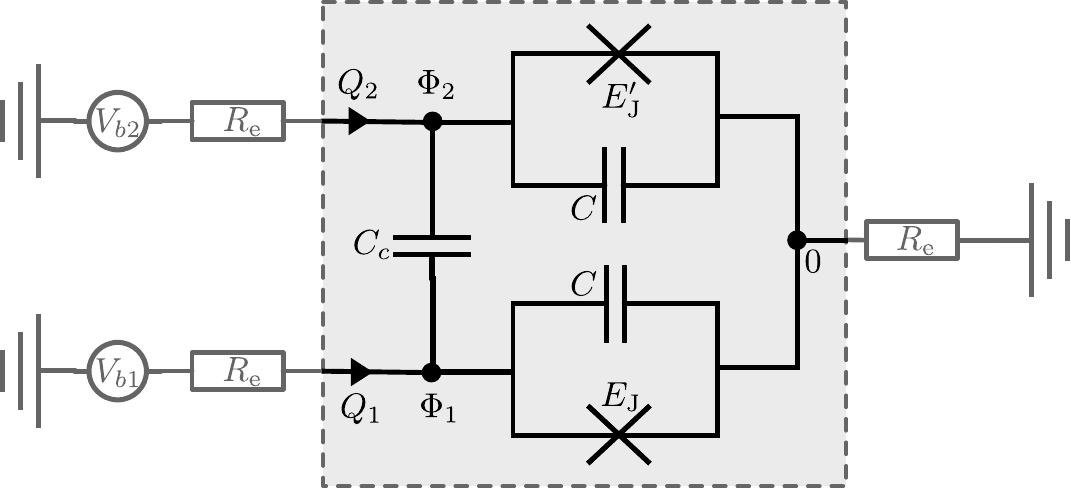}
	\caption{Undamped part (gray box) of the synchronization circuit consisting of two small Josephson junctions with a coupling capacitance $C_{\rm c}$. We model the small Josephson junctions as junctions with a shunt capacitance $C$, where $E_{\rm J}'$ corresponds to the tunable Josephson energy of the SQUID. The dynamical variables in the circuit are the node fluxes $\Phi_i$ whereas the externally supplied charges $Q_i$ are slow due to the large biasing resistors $R_\text{e}$. The circuit is supplied with DC bias voltages $V_{{\rm b}i}$.}
	\label{FigS1}
\end{figure}

This is the supplemental material for our article ``On chip synchronization of Bloch oscillations in a strongly coupled pair of small Josephson junctions''. Here, we present the details of the semi-classical circuit model used for our simulations. We derive the lowest band energy of the circuit highlighted in Fig.~\ref{FigS1} which is the undamped part of the circuit discussed in the article. 

The circuit consists of two small Josephson junctions that are coupled capacitively with one junction being a SQUID with a tunable Josephson energy $E_{\rm J}'$. We model the small junctions by attributing a shunt capacitance $C$ to both of them. Physically, this effective capacitance represents the tunnel junction capacitance, but also includes the stray contribution of the current leads. For the sake of simplicity, we choose both effective capacitances to be equal. The circuit is connected to two biasing leads that supply charges $Q_i$. Since the resistance of the biasing leads is much larger than the quantum resistance, the charges are slow variables compared to the node fluxes $\Phi_i$ that determine the dynamics of the circuit.
Using the dimensionless node fluxes $\varphi_i=(2e/\hbar)\Phi_i$, the Euclidean Lagrangian of the undamped circuit is given by $\mathcal{L}= T + U + \mathcal{L}_\text{top}$
\begin{align}
	T&=\frac{\hbar^2}{8e^2}\Bigl[C(\dot\varphi_1^2+\dot\varphi_2^2)+C_{\rm
	c}(\dot\varphi_1-\dot\varphi_2)^2\Bigr]\,, & 
	U&=E_{\rm
	J}(1-\cos\varphi_1)+E_{\rm J}'(1-\cos\varphi_2)\,, &
		\mathcal{L}_\text{top} &=-i\frac{\hbar}{2e}(\dot\varphi_1Q_1+\dot\varphi_2Q_2),
\end{align}
where for slow $Q_i$ the Aharonov-Casher term $\mathcal{L}_\text{top}$ is a total time derivative and therefore does not influence the dynamics of the node fluxes.

\begin{figure}[tb]
	\centering\includegraphics{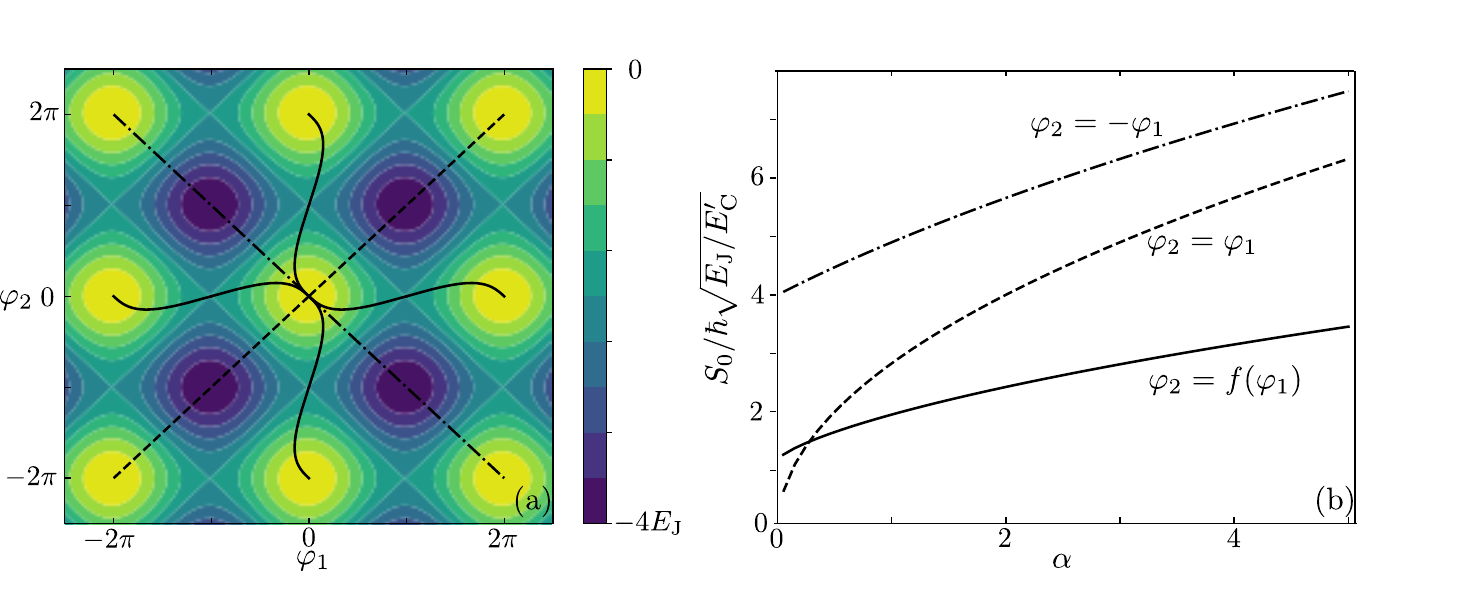}
	\caption{Relevant trajectories with their associated actions in the case of $\beta=1$. (a) Trajectories  at $\alpha=0.3$ in the potential energy landscape given by $-U$. (b) Numerical values of the actions for different $\alpha$.}
	\label{FigS2}
\end{figure}

 This Lagrangian describes a particle with anisotropic mass in a 2-dimensional cosine potential. In the WKB-approximation, the contributions to the lowest energy band of the system are determined to exponential accuracy by
 \begin{equation}
	 E_j\propto e^{-S_j/\hbar},
 \end{equation}
 where $S_j$ is the Euclidean action of a trajectory from one potential minimum to an adjacent minimum. The action of each trajectory can be split into two parts
\begin{equation}
	S_j= S_{0,j}+\Delta S_j,
\end{equation} 
where $\Delta S$ corresponds to the contribution of $\mathcal{L}_\text{top}$ which only depends on the difference $\Delta\varphi_{i,j}$ between the starting- and endpoints of the trajectory with
\begin{equation}
\Delta S_j=-i\frac{\hbar}{2e}(\Delta\varphi_{1,j} Q_1+\Delta\varphi_{2,j} Q_2).
\end{equation}
Since the Lagrangian is invariant under inversion of the node fluxes, for each trajectory there is a second inverted trajectory with the same action. 
We obtain the trajectory-dependent contribution to the action $\tilde S_j$ by considering the abbreviated action $S_{0,j}$ of a zero energy trajectory between two minima following  the Euler-Maupertuis principle. Therefore the resulting energy contribution is given by~\cite{LandauLifshitz}
\begin{equation}
	E_j\sim - \sqrt{E'_{\rm C} E_{\rm J}}e^{-S_{0,j}/\hbar}\cos\left(\frac{\Delta\varphi_{1,j}Q_1+\Delta\varphi_{2,j}Q_1}{2e}\right),
\end{equation}
with the charging energy $E'_{\rm C}=e^2/2C_{\rm c}$ of the coupling capacitance where the prefactor corresponds to the classical attempt frequency.
The action is given by
\begin{equation}
	S_0=\frac{\hbar}{4}\sqrt{\frac{E_{\rm J}}{E'_{\rm C}}}\int_{\tau_0}^{\tau_1}\Bigl\{\Bigl[1+\beta-\cos\varphi_1-\beta\cos\varphi_2\Bigr]\Bigl[\alpha(\dot\varphi_1^2+\dot\varphi_2^2)+(\dot\varphi_1-\dot\varphi_2)^2
	\Bigr]\Bigr\}^{1/2}d\tau,
\end{equation} 
with $\alpha=C/C_{\rm c}$ and $\beta=E_{\rm J}'/E_{\rm J}$ where $\tau$ parametrizes the trajectory. 
Choosing the parametrization $\varphi_1(\tau)=\tau$ and $\varphi_2(\tau)=f(\tau)$, we obtain an Euler-Lagrange equation for the trajectories with stationary action
\begin{equation}
	f''(\tau)=\frac{1+\alpha-2f'(\tau)+(1+\alpha)f'(\tau)^2}{2\alpha(2+\alpha)[\cos\tau+\beta\cos f(\tau)-1-\beta]}\left\{\Bigl[(1+\alpha)\sin\tau+\beta\sin f(\tau)\Bigr]f'(\tau)-\sin\tau-(1+\alpha)\beta\sin f(\tau)\right\}.
\end{equation}
From the invariance of the action functional under the transformation
\begin{equation}
f(\tau)\mapsto \tilde f(\tau)=-f(2\pi-\tau),
\end{equation}
we conclude that the solutions to the Euler-Lagrange equation either possess the symmetry
\begin{equation}
\label{eq:symmetry}
f(\tau)=-f(2\pi-\tau),
\end{equation}
or can be mapped onto equal trajectories with starting and end points that are shifted by multiples of $2\pi$ which reflects the periodicity of the potential. If $f(\tau)$ has a nonlinear dependence on $\tau$, it must possess the symmetry since the transformation does not correspond to a linear shift of the whole function in that case.
In the case of equal Josephson energies corresponding to $\beta=1$, we obtain two analytical solutions given by
\begin{equation}
f_{\pm}(\tau)=\pm\tau,
\end{equation}
with corresponding actions
\begin{equation}
	S_{0+}=\hbar\sqrt{\frac{8E_{\rm J}}{E'_{\rm C}}}\sqrt{\alpha}\qquad\mathrm{and}\qquad S_{0-}=\hbar\sqrt{\frac{8E_{\rm J}}{E'_{\rm C}}}\sqrt{2+\alpha}\,.
\end{equation}
The trajectories given by $f_{\pm}(\tau)$ correspond to mutual phase tunneling
processes into next to nearest neighboring minima which correspond to a
synchronization between the two junctions. Furthermore there also exist
trajectories into the neighboring minima in all directions that correspond to
the tunneling of an individual phase. Since for $\beta=1$, the problem is
symmetric under exchange of $\varphi_1$ and $\varphi_2$, we only need to
consider one of the four neighboring minima. To obtain the corresponding
trajectory, we integrate the Euler-Lagrange equation numerically from a fixed
starting point $(\tau_0,f(\tau_0))$ while varying the initial slope
$f'(\tau_0)$ in order to find a trajectory that passes through the neighboring
minimum.  To counteract the numerical instability of the equation around the
minima of the potential, we utilize the fact that the solution must exhibit
the symmetry defined in Eq.~(\ref{eq:symmetry}). A trajectory between the points $(0,0)$ and $(2\pi,0)$ has to fulfill
$f(\pi)=0$,
 in order to obey the symmetry relation. Therefore, we use the point $(\pi,0)$ as a starting point of our integration to reduce the number of unstable regions the trajectory passes through to the targeted minimum. We show the resulting trajectories as well as their corresponding actions in Fig.~\ref{FigS2}.
The numerics show that the action corresponding to $f_-$ is far higher than the action of the individual phase slips. In the regime of $\alpha <1$ corresponding to a large coupling capacitance, the action of $f_+$ becomes comparable or even smaller than the action of the individual phase slips.
 Summing up all contributions we obtain a total lowest band energy of 
\begin{equation}
	E=-E_{{\rm c}1}\cos\left(\frac{\pi Q_1}{e}\right)-E_{{\rm c}2}\cos\left(\frac{\pi Q_2}{e}\right)-E_{+}\cos\left[\frac{\pi( Q_1+ Q_2)}{e}\right]-E_{-}\cos\left[\frac{\pi( Q_1- Q_2)}{e}\right],
\end{equation}
where $E_{-}\ll E_{{\rm c}i}$ since the action of the corresponding trajectory is considerably higher. Since we consider the setup at finite temperature of $k_{\rm B} T\approx E_{{\rm c}1}$, we neglect the contribution of $E_-$ to the total energy since it is much smaller the thermal energy. Based on the energy of the lowest band, we obtain the voltage drop $U_i$ between biasing lead $i$ and the lower output according to 
\begin{equation}
U_i=\frac{\partial E}{\partial Q_i}.
\end{equation}
In the regime of maximal magnetic frustration, the Josephson energy of the SQUID becomes much smaller than the Josephson energy of the other junction. We approximate this by setting $\beta$ to zero which effectively makes the junction fully transparent. In this fully asymmetric regime, the motion in the $\varphi_2$-direction is no longer constrained. Therefore no quantization of this variable occurs which leads to a lowest band energy of
\begin{equation}
	E=-E_{{\rm c}1}\cos\left(\frac{\pi Q_1}{e}\right),
\end{equation}
since the other contributions containing $Q_1$ are now negligible as well, the effective blockade voltage in this direction is also lowered. This is expected for a transparent SQUID junction since the coupling capacitance now effectively shunts the Josephson junction leading to a lower overall charging energy which leads to a reduced critical voltage for a single junction~\cite{Arndt2018}.